\documentclass[12pt]{article}
\usepackage{a4p}
\usepackage{epsfig}
\usepackage{cite}
\usepackage{amsmath}     

\begin{document}
%
%
\newcommand{\PPEnum}    {CERN-EP/2002-050}
\newcommand{\Date}      {10 July 2002}
\newcommand{\Author}    {Shoji Asai}
\newcommand{\MailAddr}  {Shoji.Asai@@cern.ch}
\newcommand{\EdBoard}   {EB member  M.Gruwe, R.McPherson, K.Nagai and R.Stroehmer} 
\newcommand{\DraftVer}  {Version 5.0}
\newcommand{\DraftDate} {02 July. 2002}
\newcommand{\TimeLimit} {09th July. 2002; 12:00 h (CERN TIME)}
\newcommand{\Submit} {To be submitted to Phys. Lett. B}

%
 
\def\toprule{\noalign{\hrule \medskip}}
\def\midrule{\noalign{\medskip\hrule }}
\def\botrule{\noalign{\medskip\hrule }}
\setlength{\parskip}{\medskipamount}
 

\newcommand{\ee}{{\mathrm e}^+ {\mathrm e}^-}
\newcommand{\sq}{\tilde{\mathrm q}}
\newcommand{\seff}{\tilde{\mathrm f}}
\newcommand{\sele}{\tilde{\mathrm e}}
\newcommand{\sell}{\tilde{\ell}}
\newcommand{\snu}{\tilde{\nu}}
\newcommand{\ch}{\tilde{\chi}^\pm}
\newcommand{\chp}{\tilde{\chi}_{1}^+}
\newcommand{\chm}{\tilde{\chi}_{1}^-}
\newcommand{\chpm}{\tilde{\chi}_{1}^\pm}
\newcommand{\nt}{\tilde{\chi}^0}
\newcommand{\qq}{{\mathrm q}\bar{\mathrm q}}
\newcommand{\qqx}{{\mathrm q}\bar{\mathrm q'}}
\newcommand{\nunu}{\nu \bar{\nu}}
\newcommand{\mumu}{\mu^+ \mu^-}
\newcommand{\tautau}{\tau^+ \tau^-}
\newcommand{\ellell}{\ell^+ \ell^-}
\newcommand {\wenu} {{\mathrm{We}} \nu}
\newcommand{\nulqq}{\nu \ell {\mathrm q} \bar{\mathrm q}'}
\newcommand{\MZ}{M_{\mathrm Z}}
\newcommand{\MW}{M_{\mathrm W}}

\newcommand {\squark}        {\tilde{\mathrm{q}}}
\newcommand {\stopm}         {\tilde{\mathrm{t}}_{1}}
\newcommand {\stops}         {\tilde{\mathrm{t}}_{2}}
\newcommand {\stopbar}       {\bar{\tilde{\mathrm{t}}}_{1}}
\newcommand {\stopx}         {\tilde{\mathrm{t}}}
\newcommand {\sneutrino}     {\tilde{\nu}}
\newcommand {\slepton}       {\tilde{\ell}}
\newcommand {\cslepton}       {\tilde{\ell}^{\pm}}
\newcommand {\stopl}         {\tilde{\mathrm{t}}_{\mathrm L}}
\newcommand {\stopr}         {\tilde{\mathrm{t}}_{\mathrm R}}
\newcommand {\stoppair}      {\tilde{\mathrm{t}}_{1} \bar{\tilde{\mathrm{t}}}_{1}}
\newcommand {\gluino}        {\tilde{\mathrm g}}

\newcommand {\sbotm}         {\tilde{\mathrm{b}}_{1}}
\newcommand {\sbots}         {\tilde{\mathrm{b}}_{2}}
\newcommand {\sbotbar}       {\bar{\tilde{\mathrm{b}}}_{1}}
\newcommand {\sbotx}         {\tilde{\mathrm{b}}}
\newcommand {\sbotl}         {\tilde{\mathrm{b}}_{\mathrm L}}
\newcommand {\sbotr}         {\tilde{\mathrm{b}}_{\mathrm R}}
\newcommand {\sbotpair}      {\tilde{\mathrm{b}}_{1} \bar{\tilde{\mathrm{b}}}_{1}}

\newcommand {\neutralino}    {\tilde{\chi }^{0}_{1}}
\newcommand {\neutrala}      {\tilde{\chi }^{0}_{2}}
\newcommand {\neutralb}      {\tilde{\chi }^{0}_{3}}
\newcommand {\neutralc}      {\tilde{\chi }^{0}_{4}}
\newcommand {\bino}          {\tilde{\mathrm B}^{0}}
\newcommand {\wino}          {\tilde{\mathrm W}^{0}}
\newcommand {\higginoa}      {\tilde{\mathrm H_{1}}^{0}}
\newcommand {\higginob}      {\tilde{\mathrm H_{1}}^{0}}
\newcommand {\chargino}      {\tilde{\chi }^{\pm}_{1}}
\newcommand {\charginop}     {\tilde{\chi }^{+}_{1}}
\newcommand {\KK}            {{\mathrm K}^{0}-\bar{\mathrm K}^{0}}
\newcommand {\ff}            {{\mathrm f} \bar{\mathrm f}}
\newcommand {\bq}            {\mathrm b} 
\newcommand {\cq}            {\mathrm c} 
\newcommand {\ele}           {\mathrm e} 
\newcommand {\bstopm} {\mbox{$\boldmath {\tilde{\mathrm{t}}_{1}} $}}
\newcommand {\Mt}            {m_{\mathrm t}}
\newcommand {\mb}            {m_{\mathrm b}}
\newcommand {\mc}            {m_{\mathrm c}}
\newcommand {\mscalar}       {m_{0}}
\newcommand {\Mgaugino}      {M_{1/2}}
\newcommand {\tanb}          {\tan \beta}
\newcommand {\rs}            {\sqrt{s}}
\newcommand {\WW}            {{\mathrm W}^+{\mathrm W}^-}
\newcommand {\eetautau}      {\ee-\rightarrow {\tau^+}{\tau^-}}
\newcommand {\MGUT}          {M_{\mathrm{GUT}}}
\newcommand {\Zboson}        {{\mathrm Z}^{0}}
\newcommand {\Wpm}           {{\mathrm W}^{\pm}}
\newcommand {\Wp}            {{\mathrm W}^{+}}
\newcommand {\Wpoff}         {{\mathrm W}^{*+}}
\newcommand {\allqq}         {\sum_{q \neq t} q \bar{q}}
\newcommand {\mixstop}       {\theta _{\stopx}}
\newcommand {\mixsbot}       {\theta _{\sbotx}}
\newcommand {\phiacop}       {\phi _{\mathrm {acop}}}
\newcommand {\cosjet}        {\cos\thejet}
\newcommand {\costhr}        {\cos\thethr}
\newcommand {\djoin}         {d_{\mathrm{join}}}
\newcommand {\mchar}         {m_{\chpm}}
\newcommand {\mstop}         {m_{\stopm}}
\newcommand {\msbot}         {m_{\sbotm}}
\newcommand {\mchi}          {m_{\neutralino}}
\newcommand {\pp}{p \bar{p}}
 
\newcommand{\gsim}{\;\raisebox{-0.9ex}
           {$\textstyle\stackrel{\textstyle >}{\sim}$}\;}
\newcommand{\lsim}{\;\raisebox{-0.9ex}{$\textstyle\stackrel{\textstyle<}
           {\sim}$}\;}

\newcommand{\degree}    {^\circ}
%
\newcommand{\Ecm}       {E_{\mathrm{cm}}}
\newcommand{\Ebeam}     {E_{\mathrm{b}}}
\newcommand{\roots}     {\sqrt{s}}
%
%
\newcommand{\thrust}    {T}
\newcommand{\nthrust}   {\hat{n}_{\mathrm{thrust}}}
\newcommand{\thethr}    {\theta_{\,\mathrm{thrust}}}
\newcommand{\phithr}    {\phi_{\mathrm{thrust}}}
\newcommand{\acosthr}   {|\cos\thethr|}
\newcommand{\thejet}    {\theta_{\,\mathrm{jet}}}
\newcommand{\acosjet}   {|\cos\thejet|}
\newcommand{\thmiss}    { \theta_{\mathrm{miss}} }
\newcommand{\cosmiss}   {| \cos \thmiss |}
\newcommand{\pbinv}     {\mathrm{pb}^{-1}}
%
%
\newcommand{\Evis}      {E_{\mathrm{vis}}}
\newcommand{\Rvis}      {E_{\mathrm{vis}}\,/\roots}
\newcommand{\Mvis}      {M_{\mathrm{vis}}}
\newcommand{\Rbal}      {R_{\mathrm{bal}}}
\newcommand{\mjet}      {\bar{M}_{\mathrm{jet}}}
%
%
%
\newcommand{\PhysLett}  {Phys.~Lett.}
\newcommand{\PRL} {Phys.~Rev.\ Lett.}
\newcommand{\PhysRep}   {Phys.~Rep.}
\newcommand{\PhysRev}   {Phys.~Rev.}
\newcommand{\NPhys}  {Nucl.~Phys.}
\newcommand{\NIM} {Nucl.~Instr.\ Meth.}
\newcommand{\CPC} {Comp.~Phys.\ Comm.}
\newcommand{\ZPhys}  {Z.~Phys.}
\newcommand{\IEEENS} {IEEE Trans.\ Nucl.~Sci.}
%
%
\newcommand{\OPALColl}  {OPAL Collab.}
\newcommand{\JADEColl}  {JADE Collab.}
\newcommand{\etal}      {{\it et~al.}}
\newcommand{\onecol}[2] {\multicolumn{1}{#1}{#2}}
\newcommand{\ra}        {\rightarrow}   
 
 
 
 
\begin{titlepage}
%
%
\begin{center}
    \large
    EUROPEAN ORGANISATION FOR NUCLEAR RESEARCH
\end{center}
\begin{flushright}
    \large
    \PPEnum\\
    \Date
\end{flushright}
%
%
%
%
\begin{center}
    \huge\bf\boldmath
    Search for Scalar Top \\
    and Scalar Bottom Quarks at LEP\\
\end{center}
\bigskip
\bigskip
\bigskip
%
%
\begin{center}
    \LARGE
    The OPAL Collaboration \\
\end{center}
\bigskip
\bigskip
\bigskip
%
%
\begin{abstract}
Searches for a scalar top quark and a scalar bottom quark 
have been performed using a data sample of 438~pb$^{-1}$
at centre-of-mass energies of $\roots = $192 -- 209~GeV
collected with the OPAL detector at LEP.
No evidence for a signal was found.
The 95\% confidence level lower limit
on the scalar top quark mass is  
97.6~GeV if the mixing angle between the supersymmetric partners 
of the left- and right-handed states of the top quark is zero.
When the scalar top quark decouples 
from the $\Zboson$ boson, the lower limit is 95.7 GeV.
These limits were obtained assuming that
the scalar top quark decays into 
a charm quark and the lightest neutralino,
and that the mass difference between 
the scalar top quark and the lightest 
neutralino is larger than 10~GeV.
The complementary decay mode of the scalar top quark 
decaying into a bottom quark, 
a charged lepton and a scalar neutrino has also been studied.
The lower limit on the scalar top quark mass is 96.0~GeV for this decay mode,
if the mass difference between the scalar top quark and the scalar neutrino
is greater than 10~GeV and if the mixing angle of the scalar top quark is zero.
From a search for the scalar bottom quark, 
a mass limit of 96.9~GeV was obtained if the mass difference
between the scalar bottom quark and the lightest 
neutralino is larger than 10~GeV\@.
\end{abstract}
 
\bigskip
\bigskip
\begin{center}
{\large (\Submit) }\\
\bigskip

\bigskip
\end{center}
 
\bigskip
 
\end{titlepage}

\begin{center}{\Large        The OPAL Collaboration
}\end{center}\bigskip
\begin{center}{
G.\thinspace Abbiendi$^{  2}$,
C.\thinspace Ainsley$^{  5}$,
P.F.\thinspace {\AA}kesson$^{  3}$,
G.\thinspace Alexander$^{ 22}$,
J.\thinspace Allison$^{ 16}$,
P.\thinspace Amaral$^{  9}$, 
G.\thinspace Anagnostou$^{  1}$,
K.J.\thinspace Anderson$^{  9}$,
S.\thinspace Arcelli$^{  2}$,
S.\thinspace Asai$^{ 23}$
% ...
}\end{center}\bigskip
\bigskip
$^{  1}$School of Physics and Astronomy, University of Birmingham,
Birmingham B15 2TT, UK
\newline
$^{  2}$Dipartimento di Fisica dell' Universit\`a di Bologna and INFN,
I-40126 Bologna, Italy
\newline
$^{  3}$Physikalisches Institut, Universit\"at Bonn,
D-53115 Bonn, Germany
\newline
% ...
$^{  5}$Cavendish Laboratory, Cambridge CB3 0HE, UK
\newline
% ...
$^{  9}$Enrico Fermi Institute and Department of Physics,
University of Chicago, Chicago IL 60637, USA
\newline
% ...
$^{ 16}$Department of Physics, Schuster Laboratory, The University,
Manchester M13 9PL, UK
\newline
% ...
$^{ 22}$Department of Physics and Astronomy, Tel Aviv University,
Tel Aviv 69978, Israel
\newline
$^{ 23}$International Centre for Elementary Particle Physics and
Department of Physics, University of Tokyo, Tokyo 113-0033, and
Kobe University, Kobe 657-8501, Japan
\newline
% ...

\newpage

\section{Introduction}

Supersymmetric (SUSY) extensions of the Standard Model
predict the existence of bosonic partners of all known fermions.
The scalar top quark~($\stopx$), which is the bosonic partner of the 
top quark, may be light
because of supersymmetric radiative corrections~\cite{stop1}\@. 
Furthermore, the supersymmetric partners of the
right-handed and left-handed top quarks
($\stopr$ and $\stopl$) mix, and
the resulting two mass eigenstates ($\stopm$ and $\stops$)
have a mass splitting which 
may be very large due to the large top quark mass.  
The resulting lighter mass eigenstate ($\stopm$),
$\stopm = \stopl \cos \mixstop + \stopr \sin \mixstop$,
where $\mixstop$ is a mixing angle,
can be lighter than any other
charged SUSY particle, and also lighter than the top quark~\cite{stop1}\@.
All SUSY breaking parameters are absorbed 
in $\mixstop$ and the mass of $\stopm$\@.

The scalar bottom quark ($\sbotx$) 
can also be light if $\tanb$, the ratio of vacuum expectation
values of the two Higgs doublet fields, is large.
In this case, the analogous mixing between the supersymmetric partners 
of the right- and left-handed states 
of the bottom quark ($\sbotr$ and $\sbotl$) becomes large,
and the resulting two mass eigenstates ($\sbotm$ and $\sbots$)
also have a large mass splitting~\cite{bartl}\@.
The mass of the lighter mass eigenstate 
($\sbotm$) may therefore be within the reach of LEP. 

Assuming R-parity~\cite{RP} conservation and that 
the $\neutrala$ and $\cslepton$ are heavier than the $\stopm$,
the dominant decay mode of the $\stopm$ is  
expected to be either
$\stopm \ra \cq \neutralino$ or $\stopm \ra \bq \snu \ell^{+}$,
where $\neutralino$ is the lightest neutralino, $\snu$ is the scalar neutrino,
and $\ell$ is $\ele$, $\mu$ or $\tau$\@.
The latter decay mode is dominant if it is kinematically allowed.
Otherwise the flavour changing two-body decay, 
$ \stopm \ra \cq \neutralino$, is dominant
except for the small region where $m_{\stopm}-\mchi > m_{\Wpm} + \mb$
\footnote{ In this region, $\stopm \ra \bq  \neutralino \Wp$ becomes
dominant through a virtual chargino. 
This decay mode has not been studied in this paper.}\@.
Both of these decay modes ($\stopm \ra \cq \neutralino$
and $\stopm \ra \bq \snu \ell^{+}$) have been searched for. 
The dominant decay mode of the $\sbotm$ is 
expected to be $ \sbotm \ra \bq \neutralino$\@.
Since the decay widths of these modes are smaller than
the QCD energy scale, 
the $\stopm$ and $\sbotm$ produce colourless squark-hadrons
before decay.    
Under the assumption of  R-parity conservation,
$\neutralino$ and $\snu$ are invisible in the detector.
Thus, $\stoppair$ and $\sbotpair$ events 
are characterised by two acoplanar jets
\footnote{Two jets are called `acoplanar' if they
not back-to-back with each other in the plane 
perpendicular to the beam axis.}
or two acoplanar jets plus
two leptons, with missing energy.
The phenomenology of the production and decay of 
$\stopm$ and $\sbotm$ is described in Section 2 of 
Ref.~\cite{stop171}\@.

The CDF Collaboration has reported lower limit values~\cite{CDF} 
on the $\stopm$ mass of 89 and 110~GeV (95\% C.L.),
when the mass difference between $\stopm$ and $\neutralino$ 
is larger than about 40 and 60~GeV, respectively.
These limits were obtained with the assumption
that $\stopm \ra \cq \neutralino$\@.
Searches at $\ee$ colliders are sensitive to smaller mass differences. 
The first lower limits on the $\stopm$ mass were obtained around 
the $\Zboson$ peak (LEP1) assuming 
$\stopm \ra \cq \neutralino$~\cite{opalstop}\@.
Using part of the higher energy LEP2 data sample, the 95\% C.L. lower limit
for a mass difference larger than 6~GeV 
was improved to 83~GeV~\cite{stop189}\@.
Several other squark searches at various centre-of-mass energies 
($\roots$) have also been performed
at LEP~\cite{stop171, stop161, stop183, alephstop, alephstop2}\@.

For the decay mode of $\stopm \ra \bq \snu \ell^{+}$
the first lower limit on the $\stopm$ mass was obtained  
at $\roots$ = 161~GeV~\cite{stop161}, and  
successive searches were performed at
LEP~\cite{stop171, stop183, stop189, alephstop, alephstop2}
and the Tevatron.
The D0 Collaboration has reported
a lower limit~\cite{D0} on the $\stopm$ mass of 123~GeV (95\% C.L.), 
when the mass difference between $\stopm$ and $\snu$ 
is larger than 40~GeV and the branching fraction to each lepton flavour
is the same.
A search for the four-body decay mode, $\stopm \ra \bq  \neutralino \Wpoff$,
where the $\mathrm{W}$~boson is off shell,
was recently performed at LEP and no evidence was reported~\cite{alephstop2}\@.

In 1999 and 2000, the LEP $\ee$ collider at CERN operated 
at $\rs$= 192--209~GeV,
and a data sample of about 440~$\pbinv$ was collected with
the OPAL detector.
Luminosities and mean values of $\rs$ are summarised in Table~\ref{tab:lumi}\@. 
\begin{table}[h]
\centering
\begin{tabular}{|l||r|r|}
\hline
$\rs$ range & Luminosity-weighted & Luminosity  \\
(GeV) & $<\rs>$ (GeV) & (pb$^{-1}$)\\
\hline
190-194 & 191.6 & 29.1 \\
194-198 & 195.5 & 74.0 \\
198-201 & 199.5 & 75.4 \\
201-204 & 201.6 & 38.3 \\
204-206 & 204.9 & 82.0 \\
$>$206  & 206.5 & 138.8 \\ \hline
all     & 201.7 & 437.6 \\
\hline
\end{tabular}
\caption[]
{
List of luminosities and mean values of $\rs$ for data
collected in 1999 and 2000. 
}
\label{tab:lumi}
\end{table}

In this paper direct searches for $\stopm$ and $\sbotm$
using this data sample are reported.
The limits shown here have been obtained by combining the results
obtained at these new centre-of-mass energies 
with those previously obtained using the OPAL data at lower
$\rs$ ~\cite{stop171,stop161,stop183,stop189}\@. 

\section{The OPAL Detector and Event Simulation}

The OPAL detector,
which is described in detail in Ref.~\cite{OPAL-detector},
is a multipurpose apparatus
having nearly complete solid angle coverage.
The central detector consists of
a silicon strip detector and tracking chambers,
providing charged particle tracking 
for over 96\% of the full solid angle,
inside a uniform solenoidal magnetic field of 0.435~T\@.
A lead-glass electromagnetic calorimeter (ECAL)
located outside the magnet coil is hermetic in 
the polar angle range of $|\cos \theta |<0.984$\@.
The magnet return yoke consisting of barrel and endcap sections along with
pole tips is instrumented for hadron calorimetry (HCAL)
in the region $|\cos \theta |<0.99$\@.
Four layers of muon chambers cover the outside of the hadron calorimeter.
Forward detectors (FD), silicon-tungsten calorimeters (SW) and 
the gamma-catcher detectors (GC) are located in the forward region
($|\cos \theta|>0.98$) surrounding the beam pipe
and provide complete acceptance down to 25~mrad.

Monte Carlo simulation of the production and decays
of $\stopm$ and $\sbotm$ were performed following~\cite{stopgen}\@.
The squark ($\squark$) pairs were generated,
and the hadronisation process was subsequently performed
to produce colourless $\squark$-hadrons and other fragmentation products
according to the Lund string fragmentation scheme
(JETSET 7.4)~\cite{lund,fragment}\@.
The parameters for perturbative QCD and fragmentation processes
were optimised using hadronic $\Zboson$ decays
measured by OPAL~\cite{opalfragment}\@.
For the fragmentation of $\squark$, the fragmentation function proposed
by Peterson {\it et al.}~\cite{lund,Peterson} was used.
The $\squark$-hadron was formed from a squark and a spectator anti-quark
or diquark.
For the $\stopm$ decaying into $\cq \neutralino$,
a colour string was connected between the charm quark and the spectator.
The decays $\sbotm \ra \bq \neutralino$ 
and $ \stopm \ra \bq \ell^{+} \snu $  were simulated in a similar manner.
One thousand events were generated at each point of a two dimensional grid
of spacing of typically 5~GeV steps in $(m_{\stopm}, m_{\neutralino})$ 
for $\stopm \ra \cq \neutralino$, 
in $(m_{\stopm}, m_{\snu})$ for 
$\stopm \ra \bq \ell^{+} \snu$ (with equal branching ratios for
$\ele$, $\mu$ and $\tau$) and $\stopm \ra \bq \tau^{+} \snu$,
and in $(m_{\sbotm}, m_{\neutralino})$ for $\sbotm \ra \bq \neutralino$\@.
Smaller steps were used for the case of small mass differences
($\Delta m  =  \mstop - \mchi$, $ \mstop - m_{\snu}$ or
$ m_{\sbotm} - m_{\neutralino}$)\@. 
The signal samples were generated at $\rs$=192, 196, 200 and 206 GeV\@.

The background processes were simulated as follows.
The KK2f generator~\cite{kk2f} was used to 
simulate multihadronic ($\qq(\gamma)$) events,
$\tau^+ \tau^- (\gamma)$, and $\mumu (\gamma)$ events.  
Bhabha events, $\ee \ra \ee (\gamma) $, were generated with 
the BHWIDE program~\cite{BHWIDE}\@.
Two-photon processes are the most important background
for the case of small mass differences,
since in such cases signal events have small visible energy and 
small transverse momentum relative to
the beam direction.
Using the Monte Carlo generators 
PHOJET~\cite{PHOJET}, PYTHIA~\cite{lund} and HERWIG~\cite{HERWIG},
hadronic events from various two-photon processes were simulated
in which the invariant mass of the photon-photon
system ($M_{\gamma \gamma}$) was larger than 5.0~GeV\@.
Monte Carlo samples for leptonic two-photon processes
($\ee \ee$, $\ee \mumu$ and 
$\ee \tautau$) were generated with the Vermaseren 
program~\cite{Vermaseren}\@.
The grc4f~\cite{grace} and KoralW~\cite{koral4f} generators were used for 
all four-fermion processes except for regions
covered by the two-photon simulations.
All interference effects of the various diagrams
are taken into account in these generators.
Four-fermion processes in which at least one
of the fermions is a neutrino constitute a serious background
at large mass differences.
The generated signal and background events were processed
through the full simulation of the OPAL detector~\cite{GOPAL},
and the same analysis chain was applied as to the data.

\section{Analysis}
 
Since the event topologies of $\stopm \ra \cq \neutralino$ 
and $\sbotm \ra \bq \neutralino $ are very similar, 
the same selection criteria were used (Section~3.1, analysis A)\@. 
In Section 3.2 (analysis B), the selection criteria for
$\stopm \ra \bq \ell^{+} \snu$ are discussed. 
These analyses are the same as those in Ref.~\cite{stop189}.
Variables used to make the selections, such as 
the total visible energy and the total transverse momentum,
and jet properties, were calculated as follows.
First, the four-momenta of the tracks and 
those of the ECAL and HCAL clusters not
associated with charged tracks were summed.
Whenever a calorimeter cluster had associated charged tracks,
the expected energy deposited by the tracks was subtracted
from the cluster energy to reduce double counting.
If the energy of a cluster was smaller
than the expected energy deposited by the associated tracks,
the cluster energy was not used.

The following three preselections, 
which are common to analyses A and B,
were applied first: 
\begin{itemize}
\item[{\bf(1)}]
The number of charged tracks was required to be at least four 
and the visible mass of the event  was required to be larger than 3~GeV\@.
\item[{\bf(2)}]
The energy deposited had to be less than 5,2 and 5~GeV
in each side of the SW, FD and  GC detectors, respectively, 
to reduce the background from two-photon processes.
\item[{\bf(3)}]
The visible energy in the region of $|\cos \theta|>0.9$
was required to be less than 10\% of the total visible energy, 
and the polar angle of the missing momentum direction, $\thmiss$,
was also required to satisfy $\cosmiss < 0.9$
to reduce the two-photon and the $\qq(\gamma)$ background.
\end{itemize}

\subsection{Analysis A: {\boldmath$\boldsymbol{\stopm \ra \cq \neutralino}$} 
and {\boldmath$\boldsymbol{\sbotm \ra \bq \neutralino}$}} 
 
The experimental signature for 
$\stoppair$($\stopm \ra \cq \neutralino$) events
and $\sbotpair$ events is two jets which are not coplanar with the beam axis.
The fragmentation functions of $\stopm $ and $\sbotm $ are expected to be
hard and the invariant mass of the charm (or bottom) quark and 
the spectator quark is small,
therefore the jets are expected to be narrow and have low invariant masses.
The following five selections were applied.

\begin{itemize}
\item[{\bf(A1)}]
Events from two-photon
processes were largely removed by demanding that the missing transverse
momentum, 
$P_t$, is greater than 4.5~GeV\@.
Fig.~1(a) shows the distribution of $P_t$ 
after the preselection.
\item[{\bf(A2)}]
The number of reconstructed jets was required to be exactly two.
Jets were reconstructed using the Durham algorithm~\cite{DURHAM} 
with the jet resolution parameter of 
$y_{\rm cut}$ = $0.005 (\Evis / \rs)^{-1}$,
where $\Evis$ is the total visible energy.
This $\Evis$-dependent $y_{\rm cut}$ parameter
was necessary for good jet reconstruction over a wide range of 
$\mstop$, $\msbot$ and $\mchi$, 
and the distribution of the number of reconstructed jets is shown in Fig.~1(b)\@.
Both reconstructed jets were required to contain 
at least two charged particles
to reduce the $\tau^{+} \tau^{-}$ background\@.
\item[{\bf(A3)}]
The acoplanarity angle, $\phiacop$,
is defined as $\pi$ minus the azimuthal opening angle
between the directions of the two reconstructed jets.
To ensure the reliability of the calculation of $\phiacop$,
both jet axes were required to have a polar angle satisfying
$|\cos{\theta}_{\rm jet}| < 0.95$\@.
The value of $\phiacop$ was required to be larger than 20$\degree$\@.
\item[{\bf(A4)}]
Softness, $\cal{S}$, was defined as 
($\frac{{M}_1}{{E}_1} + \frac{{M}_2}{{E}_2}$),
where ${M}_1$ and ${M}_2$ are the invariant masses 
of the two reconstructed jets, 
and ${E}_1$ and ${E}_2$ are the energies of the jets.
The signal events have low values of $\cal{S}$,  
whereas two-photon events
which pass the acoplanarity cut have relatively large values~\cite{stop183}\@. 
It was required that $1.5 \times \cal{S}$ $< (P_t-4.5)$,
where $P_t$ is 
given in units of GeV\@.
\item[{\bf(A5)}]
The arithmetic mean of the invariant masses of the jets, $\mjet$, 
was required to be smaller than 8~GeV\@.
When the invariant mass of the event, $\Mvis$, was larger than 65~GeV,
a harder cut, $\mjet < $ 5~GeV, was applied to 
reduce background from $\wenu$ events.
Fig.~1(c) shows the $\mjet$ distributions for 
data, the simulated background processes and
typical $\stoppair$ events.
As shown in this figure,
jets from $\stopm$ are expected to have
low invariant masses.
\end{itemize}

The numbers of events remaining after each cut are listed
in Table~\ref{tab:nevA}\@.
The table also shows 
the corresponding numbers of simulated events for background processes.
After all cuts, 13 events were observed in the data,
which is consistent with the expected number of background events
of 19.8$\pm$2.2\@.
Fig.~1(d) shows the $\Evis$ distribution 
after all selections were applied.

\begin{table}[h]
\centering
\begin{tabular}{|l||r||r||r|r|r|r||r|r|r|}
\hline
   &  \multicolumn{1}{c||}{data} & \multicolumn{1}{c||}{total} &
\multicolumn{1}{|c|}{$\qq (\gamma)$} & \multicolumn{1}{|c|}{$\ellell (\gamma)$} & 
\multicolumn{1}{|c|}{two-} & \multicolumn{1}{c||}{4-f} &
\multicolumn{3}{c|}{efficiency for}  \\
    &    &   \multicolumn{1}{c||}{bkg.} 
&  &  & \multicolumn{1}{c|}{photon}  &  
& \multicolumn{3}{c|}{\rule{0mm}{4mm} $\stoppair$ and $\sbotpair$}  \\
\hline
$m_{\stopm}$ (GeV)&       &      &         &          &            &
        & \multicolumn{1}{c|}{95} & 
          \multicolumn{1}{c|}{95} &
          \multicolumn{1}{c|}{--}  \\
$m_{\sbotm}$ (GeV)&       &      &         &          &            &
        & \multicolumn{1}{c|}{--} & 
          \multicolumn{1}{c|}{--} &
          \multicolumn{1}{c|}{95}  \\
$m_{\neutralino}$ (GeV)&       &      &         &            &            &
        & \multicolumn{1}{c|}{90} & 
          \multicolumn{1}{c|}{75} & 
          \multicolumn{1}{c|}{75}  \\
\hline
 cut (A1)
&  9279 &  9429 & 4147 &  1318 &  313 & 3650 & 0.40 & 0.70 & 0.70 \\
\hline
 cut (A2)
&  2189 &  2243 & 1719 &  92.0 &  112 &  319 & 0.25 & 0.62 & 0.65 \\
\hline
 cut (A3)
&   205 &   241 & 0.35 &  0.52 & 20.6 &  220 & 0.24 & 0.56 & 0.60 \\
\hline
 cut (A4)
&   198 &   229 & 0.32 &  0.52 &  9.0 &  220 & 0.18 & 0.56 & 0.59 \\
\hline
 cut (A5)
&    13 &  19.8 & 0.25 &  0.25 &  7.7 & 11.6 & 0.18 & 0.55 & 0.58 \\
& & ($\pm$2.2) & ($\pm$0.13 ) & ($\pm$0.08) & ($\pm$2.1) & 
($\pm$0.6) & & & \\
\hline
\end{tabular}
\caption[]
{
Numbers of events remaining after each cut 
for various background processes are compared with data for analysis A\@.
The simulated background processes were normalised
to the integrated luminosity of the data.
The errors due to Monte Carlo statistics are also shown.  
Efficiencies for three simulated event samples ($\roots$ = 206~GeV)
of $\stoppair$ and $\sbotpair$ are also given.
}
\label{tab:nevA}
\end{table}

The efficiencies for both $\stoppair$ and $\sbotpair $ events
are 30--60\%
if the mass difference between $\stopm$($\sbotm$) 
and $\neutralino $ is larger than 10~GeV\@.
A modest efficiency of about 20\% is obtained for 
a mass difference of 5~GeV for $\stoppair$ events.
An additional loss of 3\% (relative) arises from beam-related
background in SW, FD and GC, which was estimated using random beam crossing events.

\subsection{Analysis B: {\boldmath$\boldsymbol{\stopm \ra \bq \ell \snu}$}} 
 
The experimental signature for $\stoppair$($\stopm \ra \bq \ell \snu$) events
is two acoplanar jets plus two leptons with missing transverse momentum.
The momenta of the leptons and the missing transverse momentum
depend strongly on the mass difference between $\stopm$ and $\snu$\@.
To obtain optimal performance, 
two sets of selection criteria (analyses B-L and B-H) 
were applied for small and large mass differences, respectively.

The numbers of events remaining after each cut are listed
in Tables~\ref{tab:nevBL} and \ref{tab:nevBH}\@.
The tables also show 
the corresponding numbers for the simulated background processes.

\subsubsection{Small mass difference case} 

For the case of a small mass difference ($\Delta m \leq$ 10~GeV), 
the following four selection criteria were applied.
Lepton identification was not used in this analysis.

\begin{itemize}
\item[{\bf(B-L1)}]
The event missing transverse momentum, $P_t$,
was required to be greater than 5~GeV\@.
\item[{\bf(B-L2)}]
The number of charged tracks was required to be at least six,
and the number of reconstructed jets was 
required to be at least four,
since the signal would contain two hadronic jets 
plus two isolated leptons.
Jets were reconstructed using the Durham algorithm~\cite{DURHAM} with 
the jet resolution parameter $y_{\mathrm{cut}}$ = 0.004\@.
Figure~2(a) shows the distribution of the number of 
reconstructed jets for 
the data, the simulated background processes and
typical $\stoppair$ events.
\item[{\bf(B-L3)}]
To examine the acoplanarity of the  remaining events,
the whole event was reconstructed as two jets using the Durham algorithm.
To ensure a good measurement of the acoplanarity angle,
$|\cos{\theta}_{\rm jet}| < 0.95$ was required for 
both reconstructed jets.
Finally, the acoplanarity angle, $\phiacop$, between these two jets
was required to be greater than 15$\degree$\@.
Fig.~2(b) shows the $\phiacop$ distributions.
\item[{\bf(B-L4)}]
The total visible energy, $\Evis$,
was required to be smaller than 60~GeV to reject
four-fermion events. 
As shown in Fig.~2(c), a large fraction of four-fermion events
are removed.
\end{itemize}

\begin{table}[h]
\centering
\begin{tabular}{|l||r||r||r|r|r|r||r|r|}
\hline
   &  \multicolumn{1}{c||}{data} & \multicolumn{1}{c||}{total} &
\multicolumn{1}{|c|}{$\qq (\gamma)$} & \multicolumn{1}{|c|}{$\ellell (\gamma)$} & 
\multicolumn{1}{|c|}{two-} & \multicolumn{1}{c||}{4-f} &
\multicolumn{2}{c|}{efficiency}  \\
    &    &   \multicolumn{1}{c||}{bkg.} 
&  &  & \multicolumn{1}{c|}{photon}  &  
& \multicolumn{2}{c|}{\rule{0mm}{4mm} for $\stoppair$}  \\
\hline 
$m_{\stopm}$ (GeV)&       &      &         &          &            &
        & \multicolumn{1}{c|}{95} & 
          \multicolumn{1}{c|}{95} \\
$m_{\snu}$ (GeV)&       &      &         &            &            &
        & \multicolumn{1}{c|}{88} & 
          \multicolumn{1}{c|}{85} \\
\hline 
 cut (B-L1)
& 8922 & 8983 & 3916 & 1274 &  230 & 3563 & 0.14 & 0.47 \\
\hline
 cut (B-L2)
& 2259 & 2252 &  560 & 0.13 & 15.6 & 1676 & 0.11 & 0.42 \\
\hline
 cut (B-L3)
&  513 &  496 & 17.6 & 0.02 & 3.73 &  474 & 0.11 & 0.39 \\
\hline
 cut (B-L4)
&    5 & 5.02 & 0.17 & 0.00 & 3.62 & 1.22 & 0.11 & 0.39 \\
& & ($\pm$1.36) & ($\pm$0.09) &  & ($\pm$1.34) & 
($\pm$0.19) & &  \\
\hline
\end{tabular}
\caption[]{
Numbers of events remaining after each cut 
for various background processes are compared with data for analysis B-L\@.
The simulated background processes
were normalised to the integrated luminosity of the data.
The errors due to Monte Carlo statistics are also shown.  
Efficiencies for
two simulated samples of $\stoppair$ are 
also given.
In these samples, produced at $\roots$ = 206~GeV,
the branching fractions to each lepton flavour
are assumed to be the same.
}
\label{tab:nevBL}
\end{table}

Five events were observed in the data after all the cuts, 
which is consistent with the number of expected background
events (5.0$\pm$1.4), mainly from two-photon processes.
The detection efficiencies are 30--40\%
if the mass difference between $\stopm$ and $\snu$
is 10~GeV, 
and if the branching fraction to each lepton flavour
is the same.
Even if the branching fraction into $\bq \tau ^{+} \snu_{\tau} $
is 100\%, the efficiencies are 25--35\%.

\subsubsection{Large mass difference case} 

The selection criteria for a large mass difference ($\Delta m > $ 10~GeV) 
are as follows:

\begin{itemize}
\item[{\bf(B-H1)}]
The event missing transverse momentum, $P_t$,
was required to be greater than 6~GeV\@.
\item[{\bf(B-H2)}]
The number of charged tracks was required to be at least six, 
and the number of reconstructed jets was required 
to be at least three.
Jets were reconstructed with 
the same jet resolution parameter ($y_{\mathrm{cut}}$ = 0.004) as in (B-L2)\@.
\item[{\bf(B-H3)}]
The same selection as (B-L3) was applied on the $\phiacop$ variable
to reject $\qq(\gamma)$ events. 
\item[{\bf(B-H4)}]
A candidate event was required to contain at least one lepton,
since a signal event would contain two isolated leptons.
The selection criteria for leptons are given in Ref.~\cite{stop183}\@.
\item[{\bf(B-H5)}]
The invariant mass of the event excluding the most energetic lepton,
$M_{\mathrm{hadron}}$,
was required to be smaller than 60~GeV
in order to reject $\WW \ra \nulqq $ events. 
As shown in Fig.~3(a), a large fraction of four-fermion
events was rejected using this requirement.
Furthermore the invariant mass excluding all identified leptons
was required to be smaller than 40~GeV\@.
\item[{\bf(B-H6)}]
Finally, the visible mass of the event, $\Mvis$, must be smaller than 80~GeV 
to reduce $\WW$ background events in which 
one of $\Wpm$'s decays into $\tau \nu$ and the other 
into ${\rm q \bar{q}^{'}(g)}$\@.
If one jet from ${\rm q \bar{q}^{'}(g)}$ was misidentified as a tau lepton,
this event could pass through the previous cut (B-H5).
Fig.~3(b) shows the $\Mvis$ distributions.
\end{itemize}

\begin{table}[h]
\centering
\begin{tabular}{|l||r||r||r|r|r|r||r|r|r|}
\hline
  &  \multicolumn{1}{c||}{data} & \multicolumn{1}{c||}{total} &
\multicolumn{1}{|c|}{$\qq (\gamma)$} & \multicolumn{1}{|c|}{$\ellell (\gamma)$} & 
\multicolumn{1}{|c|}{two-} & \multicolumn{1}{c||}{4-f} &
\multicolumn{3}{c|}{efficiency for}  \\
    &    &   \multicolumn{1}{c||}{bkg.} 
&  &  & \multicolumn{1}{c|}{photon}  &  
& \multicolumn{3}{c|}{\rule{0mm}{4mm} $\stoppair$}  \\
\hline 
$m_{\stopm}$ (GeV)&       &      &         &          &            &
        & \multicolumn{1}{c|}{90} & 
          \multicolumn{1}{c|}{90} &
          \multicolumn{1}{c|}{90}  \\
$m_{\snu}$ (GeV)&       &      &         &            &            &
        & \multicolumn{1}{c|}{80} & 
          \multicolumn{1}{c|}{70} & 
          \multicolumn{1}{c|}{45}  \\
\hline 
 cut (B-H1)
&  8241 & 8230 & 3496 & 1206 &  136 & 3393 & 0.37 & 0.65 & 0.62 \\
\hline
 cut (B-H2)
&  5138 & 5259 & 2145 & 7.08 & 27.4 & 3079 & 0.37 & 0.65 & 0.62 \\
\hline
 cut (B-H3)
&  1477 & 1534 & 63.5 & 1.72 & 5.30 & 1464 & 0.35 & 0.60 & 0.54 \\
\hline
 cut (B-H4)
&  1093 & 1172 & 28.3 & 1.37 & 2.09 & 1141 & 0.30 & 0.56 & 0.52 \\
\hline
 cut (B-H5)
&     9 & 11.0 & 0.10 & 0.31 & 1.93 & 8.78 & 0.30 & 0.56 & 0.40 \\
\hline
 cut (B-H6)
&     7 & 6.34 & 0.10 & 0.15 & 1.93 & 4.15 & 0.30 & 0.56 & 0.37 \\
& & ($\pm$1.1) & ($\pm$0.06) & ($\pm$0.06) & 
($\pm$1.0) & ($\pm$0.3) & & & \\
\hline
\end{tabular}
\caption[]{
Numbers of events remaining after each cut
for various background processes are compared with data for analysis B-H\@.
The simulated background processes were normalised
to the integrated luminosity of the data.
The errors due to Monte Carlo statistics are also shown.  
Efficiencies for
three simulated samples of $\stoppair$ are 
also given.
In these samples, produced at $\roots$ = 206~GeV,
the branching fractions to each lepton flavour
are assumed to be the same.
}
\label{tab:nevBH}
\end{table}
 
Seven candidate events were observed in the data,
which is consistent with the number of expected background
events~($6.3\pm1.1$)\@.
The dominant background arises from four-fermion processes.
The detection efficiencies are 30--60\%,
if the mass difference between the $\stopm$ 
and $\snu$ is 10~GeV, and if the $\snu$ is heavier than 30~GeV\@.
The detection efficiencies for $\stopm$ were found to be 
slightly smaller for the case where it decays purely into
$\bq \tau^{+} \snu_{\tau}$ than for the case where the branching 
fraction to each lepton flavour is assumed to be the same.
 
\section{Results}

The observed number of candidate events in each case is consistent with
the expected number of background processes.
Since no evidence for $\stoppair$ and $\sbotpair$ pair-production 
has been observed,
lower limits on $\mstop$ and $\msbot$ are calculated.
The results shown here have been obtained by combining the results
obtained at these new centre-of-mass energies 
with those previously obtained using the OPAL data at lower
$\rs$~\cite{stop171,stop161,stop183,stop189}\@. 

The systematic errors on the expected number 
of signal and background events 
were estimated in the same manner as
in the previous paper~\cite{stop183}\@.
The main sources of systematic errors on the signal 
are uncertainties in the $\stopm$ and $\sbotm$ fragmentation (5--15\%)
and in Fermi motion of the spectator quark (3--10\%)\@.
The main sources of systematic errors on the background 
are uncertainties in the generation of four-fermion processes (5\%)\@.
The background from four-fermion processes evaluated
with the grc4f and KoralW generators agreed within
the statistical error, but the small
difference was conservatively taken as a systematic error.
The limited statistics of the two-photon Monte Carlo samples also give
rise to a sizable systematic error.
Detailed descriptions are given in Ref.~\cite{stop183}\@.
Systematic errors are taken into account
when calculating limits~\cite{SYSTEM}\@.

Figure~4(a) shows the 95\% C.L. excluded regions
in the ($\mstop$ , $\mchi$) plane for $\stopm \ra \cq \neutralino$\@.
In this figure there is a triangular region of
$m_{\stopm}-\mchi > m_{\Wpm} + \mb$,
in which $\stopm \ra \bq  \neutralino \Wp$(on shell)
through a virtual chargino becomes dominant
even if the chargino is heavy.
This region is not excluded.

Figures~5(a) and (b) show the 95\% C.L. excluded regions
in the ($\mstop$ , $m_{\snu}$) plane for 
$\stopm \ra \bq \ell \snu$ ($\ell$= e,$\mu$,$\tau$)
and $\stopm \ra \bq \tau ^{+} \snu_{\tau} $, respectively.
The branching fraction to each lepton flavour $\ell^{+}$ depends
on the composition of the lightest chargino~\cite{stop171}.
As the chargino becomes more Higgsino-like, the branching fraction into 
$\bq \tau ^{+} \snu_{\tau} $ becomes large.
In the limit that the chargino is a pure Wino state,
the branching fraction to each lepton flavour is the same.
Two extreme cases in which
the branching fraction to each lepton flavour is the same, or
the branching fraction into $\bq \tau ^{+} \snu_{\tau} $ is 100\%, 
were considered here\@.

The 95\% C.L. mass bounds of $\stopm$ are listed in Table~\ref{limit1}
for two values of $\mixstop$\@.
Assuming that $\stopm$ decays into $\cq \neutralino$,
and the mass difference between $\stopm$ and $\neutralino$
is greater than 10~GeV,
$\stopm$ is found to be heavier than 97.6~GeV for $\mixstop$ = 0.0\@. 
A lower limit of 95.7~GeV is obtained
even if $\stopm$ decouples from the $\Zboson$ boson ($\mixstop$=0.98~rad),
which approximately minimizes the cross-section.
When $\stopm$ decays into $\bq \ell \snu$,
the lower limit on $\mstop$ is 96.0~GeV for the zero mixing angle case, 
assuming that the mass difference between $\stopm$ and $\snu$
is greater than 10~GeV
and that the branching fraction to each lepton flavour is the same.

\begin{table}[h] \centering
\begin{tabular}{|r || c | c || c | c | } \hline
\multicolumn{5}{|c|}{ \rule{0mm}{6mm} 
            Lower limit on $\mstop$ (GeV) } \\ \hline 
 & \multicolumn{2}{|c||}{ \rule{0mm}{6mm} $\stopm \ra \cq \neutralino$ }
 & \multicolumn{1}{|c|}{$\stopm \ra \bq \ell \snu$} 
 & \multicolumn{1}{|c|}{$\stopm \ra \bq \tau \snu_{\tau}$} \\ 
 & \multicolumn{2}{|c||}{ } 
 & \multicolumn{1}{|c|}{$\ell = {\mathrm e}, \mu, \tau$} 
 & \multicolumn{1}{|c|}{Br = 100\%} \\ \hline
 
$\mixstop$ (rad) & $ \Delta m \geq 5$~GeV & 
                 $ \Delta m \geq 10$~GeV  & $ \Delta m \geq 10$~GeV 
                 & $ \Delta m \geq 10$~GeV \\ \hline
0.0   &                           95.2 &  97.6 & 96.0 & 95.5  \\ \hline
0.98  &                           91.4 &  95.7 & 92.6 & 91.5  \\ \hline
\end{tabular}
\caption[]{
The excluded $\mstop$ region at 95\% C.L. ($\Delta m  =  \mstop - \mchi$
 or $ \mstop - m_{\snu}$)\@. 
}
\label{limit1}
\end{table}

The 95\% C.L. excluded regions in the ($m_{\sbotm}$, $m_{\neutralino}$) 
plane are shown in Fig.~4(b) for two cases $\mixsbot$= 0 and 1.17~rad\@. 
The numerical mass bounds are listed in Table~\ref{limit2}
for two values of $\mixsbot$\@.
The lower limit on the $\sbotm$-mass is found to be 96.9~GeV,
if $\Delta m$ is greater than 10~GeV and $\mixsbot$ = 0.0\@.
If the $\sbotm$ decouples from the $\Zboson$ boson ($\mixsbot$=1.17~rad), 
the lower limit is 85.1~GeV\@.
Since the electromagnetic charge of $\sbotm$ is half that of $\stopm$,
the coupling between $\gamma$ and $\sbotm$ is weaker
than between $\gamma$ and $\stopm$\@.
Therefore the production cross-section of $\sbotpair$ is strongly suppressed
when the $\sbotm$ decouples from the $\Zboson$ boson.

\begin{table}[h] \centering
\begin{tabular}{|r || c | c |} \hline
\multicolumn{3}{|c|}{ \rule{0mm}{6mm} Lower limit on $m_{\sbotm}$ (GeV)
($\sbotm \ra \bq \neutralino$) } \\ \hline
 
$\mixsbot$ (rad) & $ \Delta m \geq 7$~GeV &  $\Delta m \geq 10$~GeV \\ \hline
0.0              &               93.5      &      96.9              \\ \hline
1.17                     &       82.6      &      85.1              \\ \hline
\end{tabular}
\caption[]{
The excluded $\msbot$ region at 95\% C.L. 
($\Delta m  = m_{\sbotm} - m_{\neutralino}$)
} 
\label{limit2}
\end{table}
 
\section{Summary and Conclusion}
 
A data sample of 437.6~pb$^{-1}$ collected using the OPAL detector
at $\roots = $192--209~GeV has been analysed
to search for pair production of the scalar top quark and the 
scalar bottom quark predicted by supersymmetric theories,
assuming R-parity conservation.
No evidence was found above the background level expected 
from the Standard Model.

The 95\% C.L. lower limit on the scalar top quark mass is
97.6~GeV  
if the mixing angle of the scalar top quark is zero\@.
Even if the $\stopm$ decouples from the $\Zboson$ boson, 
a lower limit of 95.7~GeV is obtained.
These limits were estimated assuming that
the scalar top quark decays into a charm quark and the lightest neutralino 
and that the mass difference between the scalar top and the lightest 
neutralino is larger than 10~GeV\@.

Assuming a relatively light scalar neutrino
($ m_{\snu} \leq \, \mstop - m_{\bq}$), 
the complementary decay mode, in which the scalar top quark decays  
into a bottom quark, a charged lepton 
and a scalar neutrino, has also been studied. 
If the mass difference between the scalar top quark and the scalar neutrino
is greater than 10~GeV
and if the mixing angle of the scalar top quark is zero,
the 95\% C.L. lower limit on the scalar top quark mass is 96.0~GeV\@.
This limit is obtained assuming that the branching fraction to
each lepton flavour is the same.
If the branching fraction to the tau lepton is 100\%,
a lower limit of 95.5~GeV is obtained. 

The lower limit on the light scalar bottom quark mass
is found to be 96.9~GeV,
assuming that the mass difference between the scalar bottom quark and 
the lightest neutralino is greater than 10~GeV
and that the mixing angle of the scalar bottom quark is zero.
When the scalar bottom quark decouples from the $\Zboson$ boson,
a lower limit of 85.1~GeV is obtained. 
These limits are significantly improved with respect to the previous
OPAL results~\cite{stop189}, and 
are the best limits published to date.
 
\section*{Acknowledgements}

We particularly wish to thank the SL Division for the efficient operation
of the LEP accelerator at all energies
 and for their close cooperation with
our experimental group.  In addition to the support staff at our own
institutions we are pleased to acknowledge the  \\
Department of Energy, USA, \\
National Science Foundation, USA, \\
Particle Physics and Astronomy Research Council, UK, \\
Natural Sciences and Engineering Research Council, Canada, \\
Israel Science Foundation, administered by the Israel
Academy of Science and Humanities, \\
Benoziyo Center for High Energy Physics,\\
Japanese Ministry of Education, Culture, Sports, Science and
Technology (MEXT) and a grant under the MEXT International
Science Research Program,\\
Japanese Society for the Promotion of Science (JSPS),\\
German Israeli Bi-national Science Foundation (GIF), \\
Bundesministerium f\"ur Bildung und Forschung, Germany, \\
National Research Council of Canada, \\
Hungarian Foundation for Scientific Research, OTKA T-029328, 
and T-038240,\\
Fund for Scientific Research, Flanders, F.W.O.-Vlaanderen, Belgium.\\


\newpage 
\begin{figure}[t]
\vspace*{-15.mm}
\begin{center}\mbox{
\epsfig{file=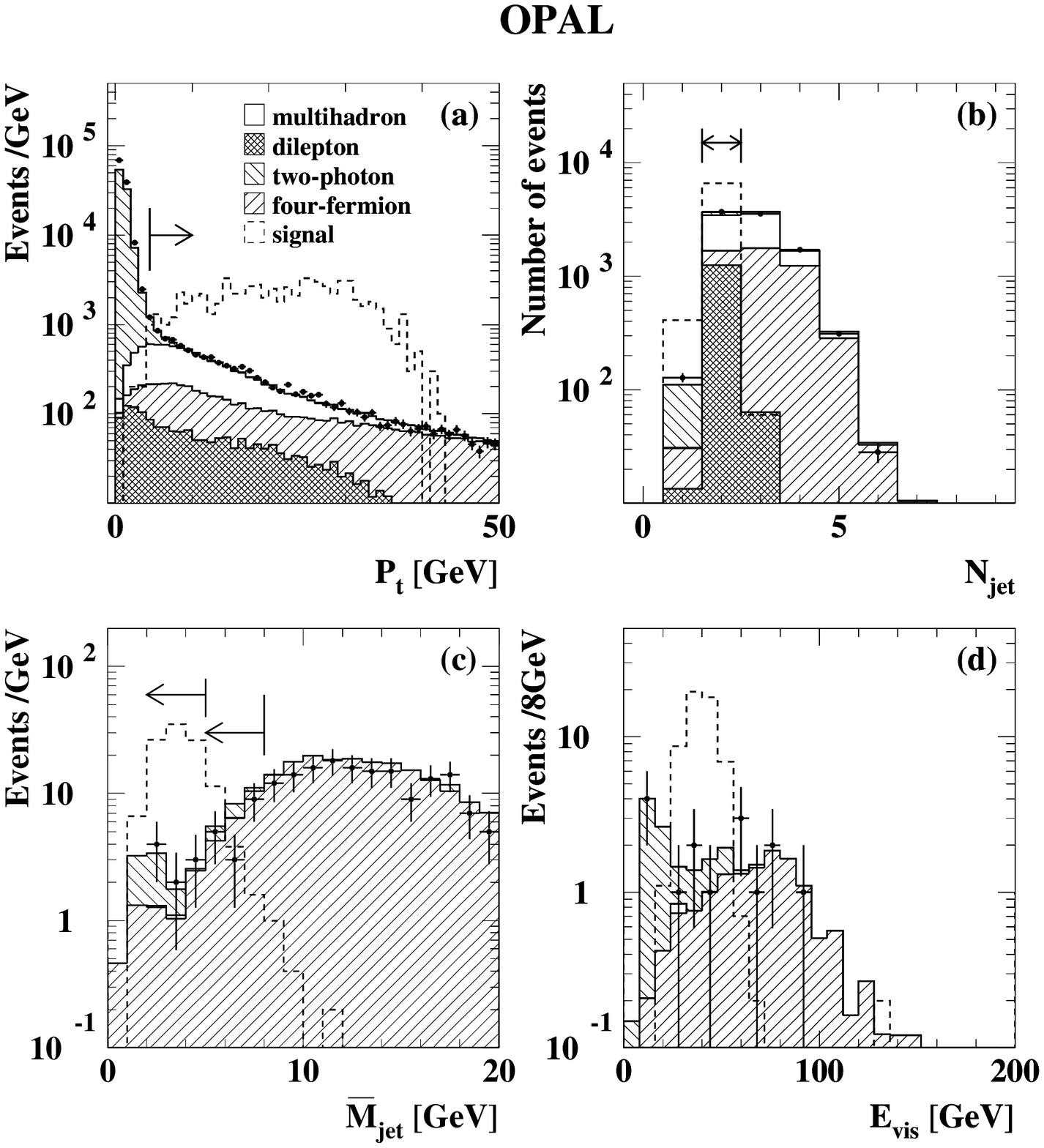,width=18.0cm}
}\end{center}
\vspace*{-3.mm}
\caption[]
{
Distributions of 
(a) ${P}_{t}$ before cut (A1), 
(b) number of reconstructed jets before cut (A2), 
(c) $\mjet$ before cut (A5), 
(d) $\Evis$ after all selections, 
for the data, simulated background events and 
typical $\stoppair$ predictions.
In these figures, the distribution of the data is shown as points 
with error bars.
The background processes are as follows: 
dilepton events (cross-hatched area), 
two-photon processes (negative slope hatched area), 
four-fermion processes (positive slope hatched area),
and multihadronic events (open area)\@.
The arrows show the cut positions.
In (c), the left (right) arrow indicates the cut position for
$\Mvis>$65~GeV ($\Mvis<$65~GeV)\@.
The predictions for $\stoppair$ signals ($\mstop$=95~GeV, $m_{\neutralino}$=75~GeV)
are shown by the dashed lines,
and the normalisations of the $\stoppair$ predictions are arbitrary.
}
\label{fig:exfig1}
\end{figure}
\newpage 
\begin{figure}[t]
\vspace*{-15.mm}
\begin{center}\mbox{
\epsfig{file=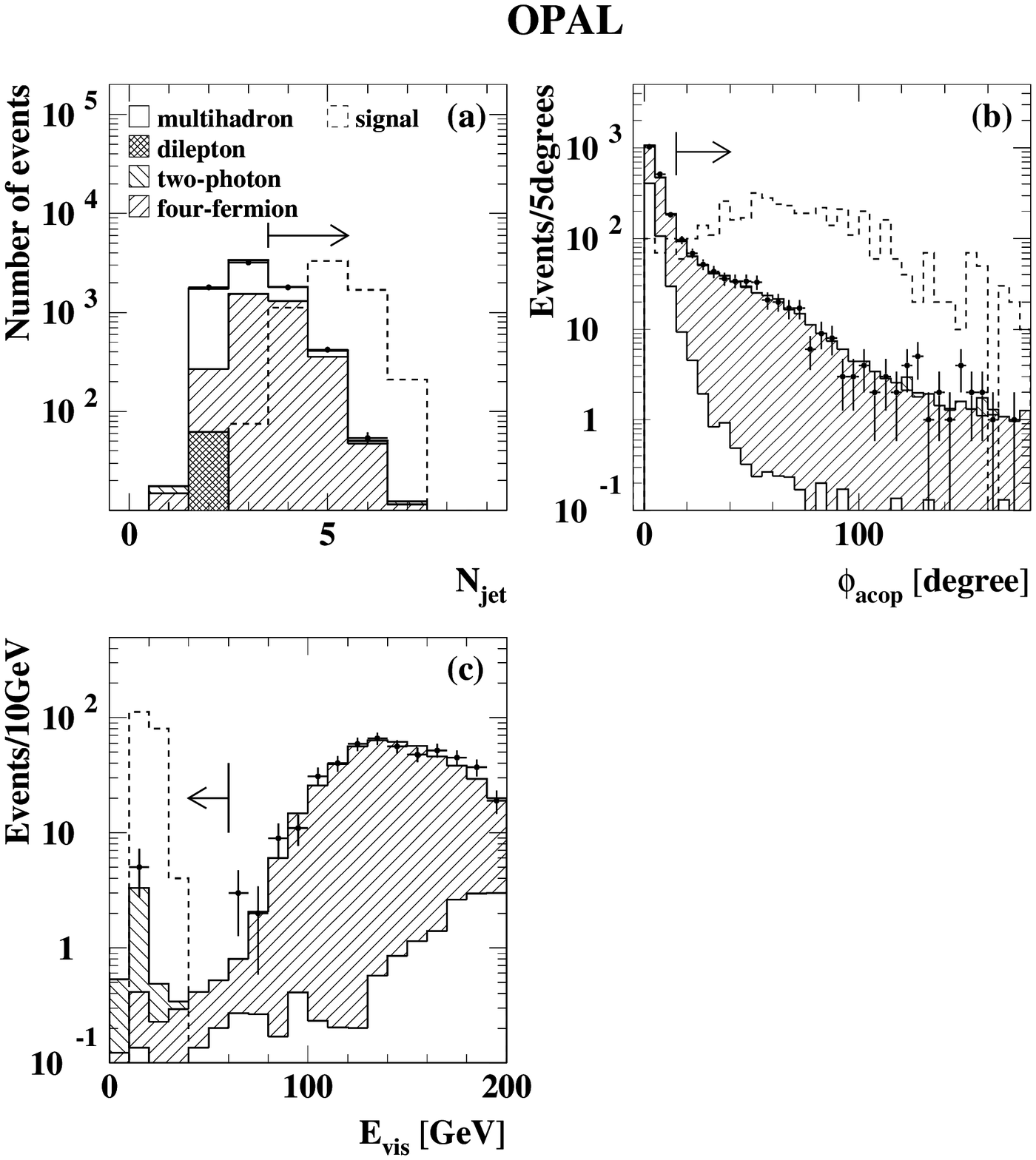,width=18.0cm}
}\end{center}
\vspace*{-3.mm}
\caption[]
{
Distributions of 
(a) number of reconstructed jets after cut (B-L1),
(b) $\phiacop$ before cut (B-L3), 
(c) $\Evis$ before cut (B-L4).
The conventions for the various histograms are the same as in Fig. 1\@.
The $\stoppair$ predictions show the cases of  
($\mstop$, $m_{\neutralino}$)=(95~GeV, 85~GeV)\@.
}
\label{fig:exfig2}
\end{figure}
\newpage 
\begin{figure}[t]
\vspace*{-15.mm}
\begin{center}\mbox{
\epsfig{file=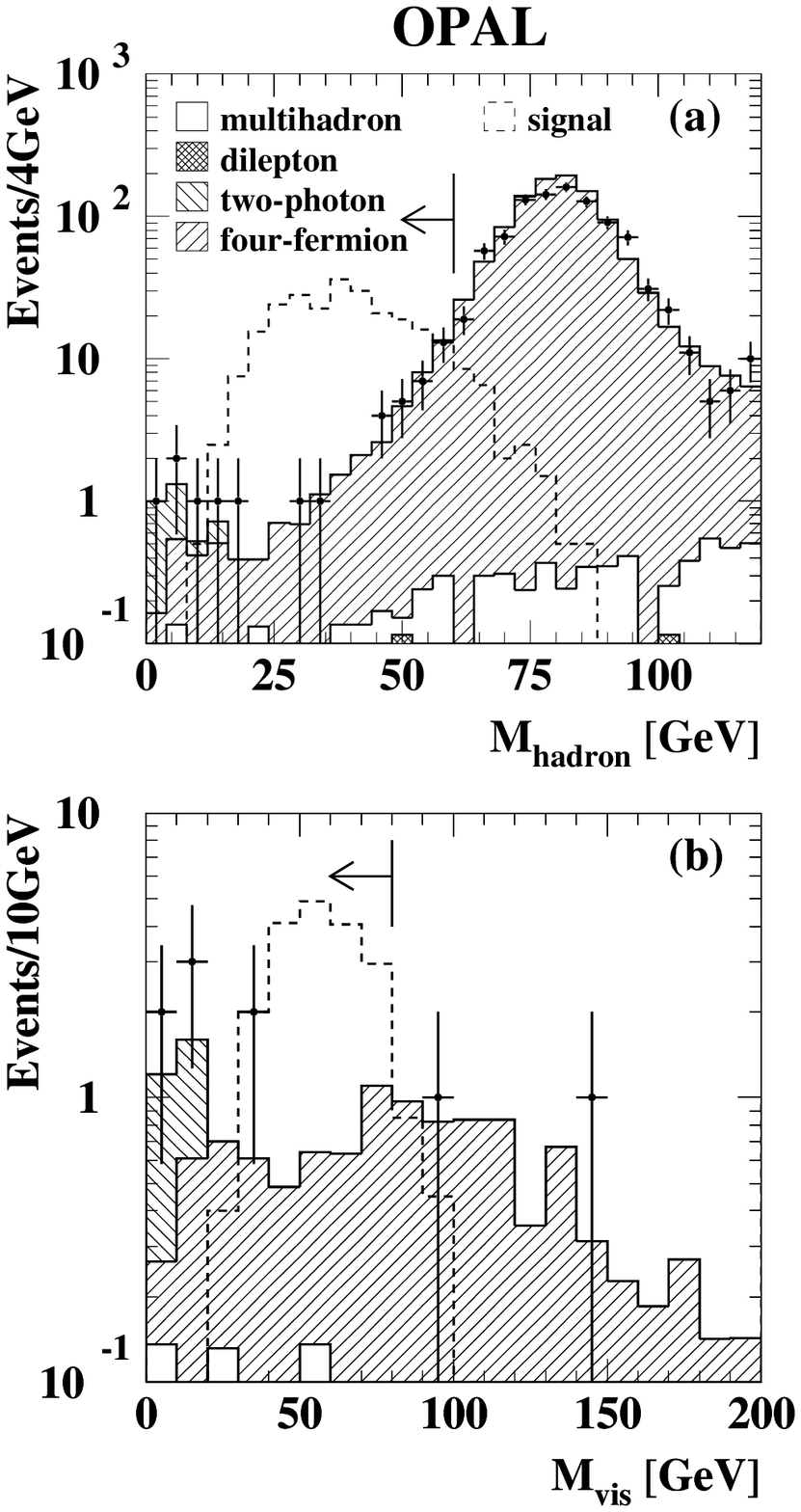,width=12.0cm}
}\end{center}
\vspace*{-3.mm}
\caption[]
{
Distributions of 
(a) invariant mass excluding the most energetic lepton before cut (B-H5),
(b)  $\Mvis$ before cut (B-H6),
The conventions for the various histograms are the same as in Fig. 1\@.
The $\stoppair$ predictions show the cases of
($\mstop$, $m_{\neutralino}$)=(95~GeV, 47.5~GeV)\@.
}
\label{fig:exfig3}
\end{figure}
\newpage 
\begin{figure}[htb]
\vspace*{-20.0mm}
\begin{center}\mbox{
\epsfig{file=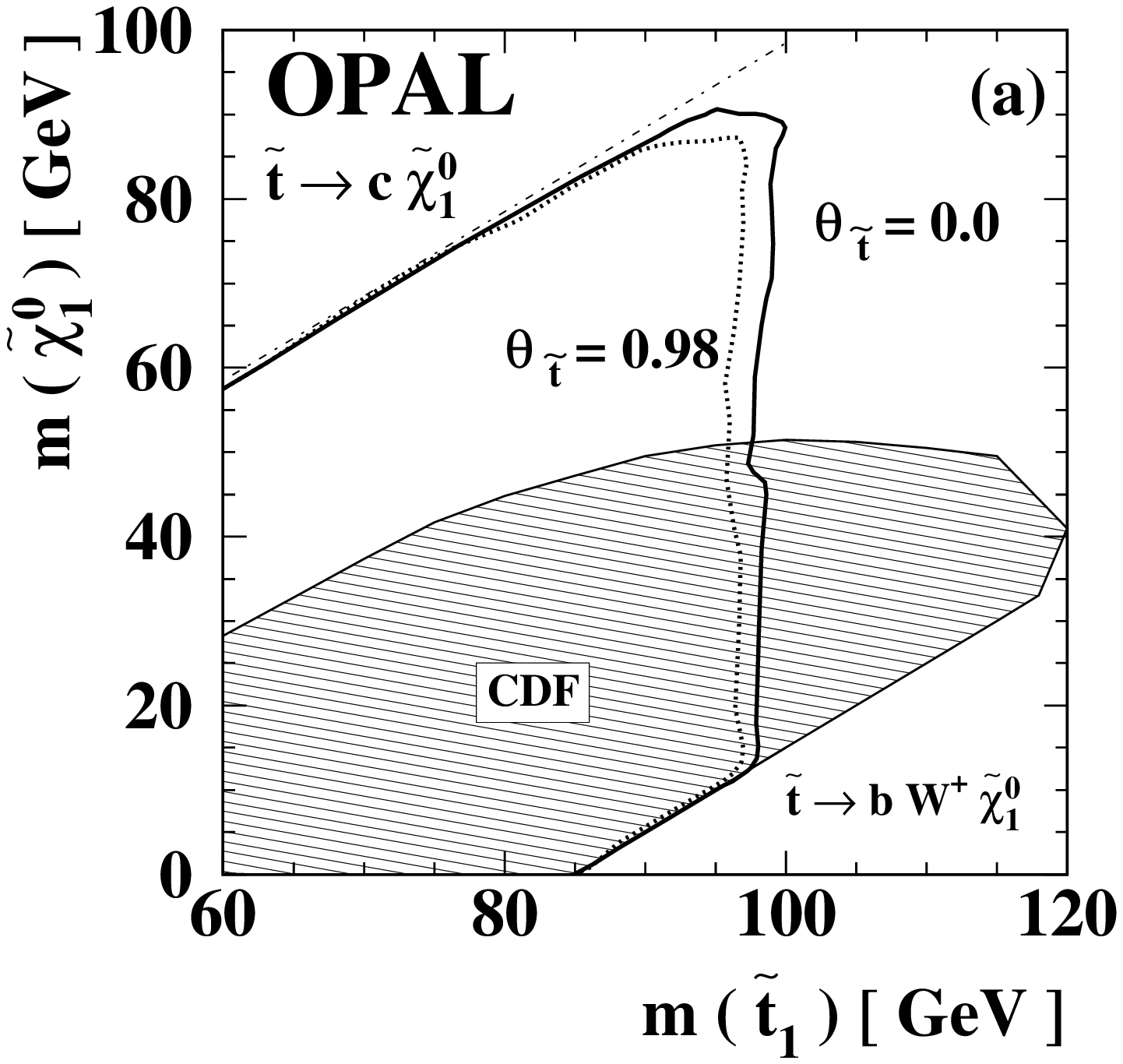,width=10.0cm} 
}\end{center} 
\vspace*{-15.0mm}
\begin{center}\mbox{
\epsfig{file=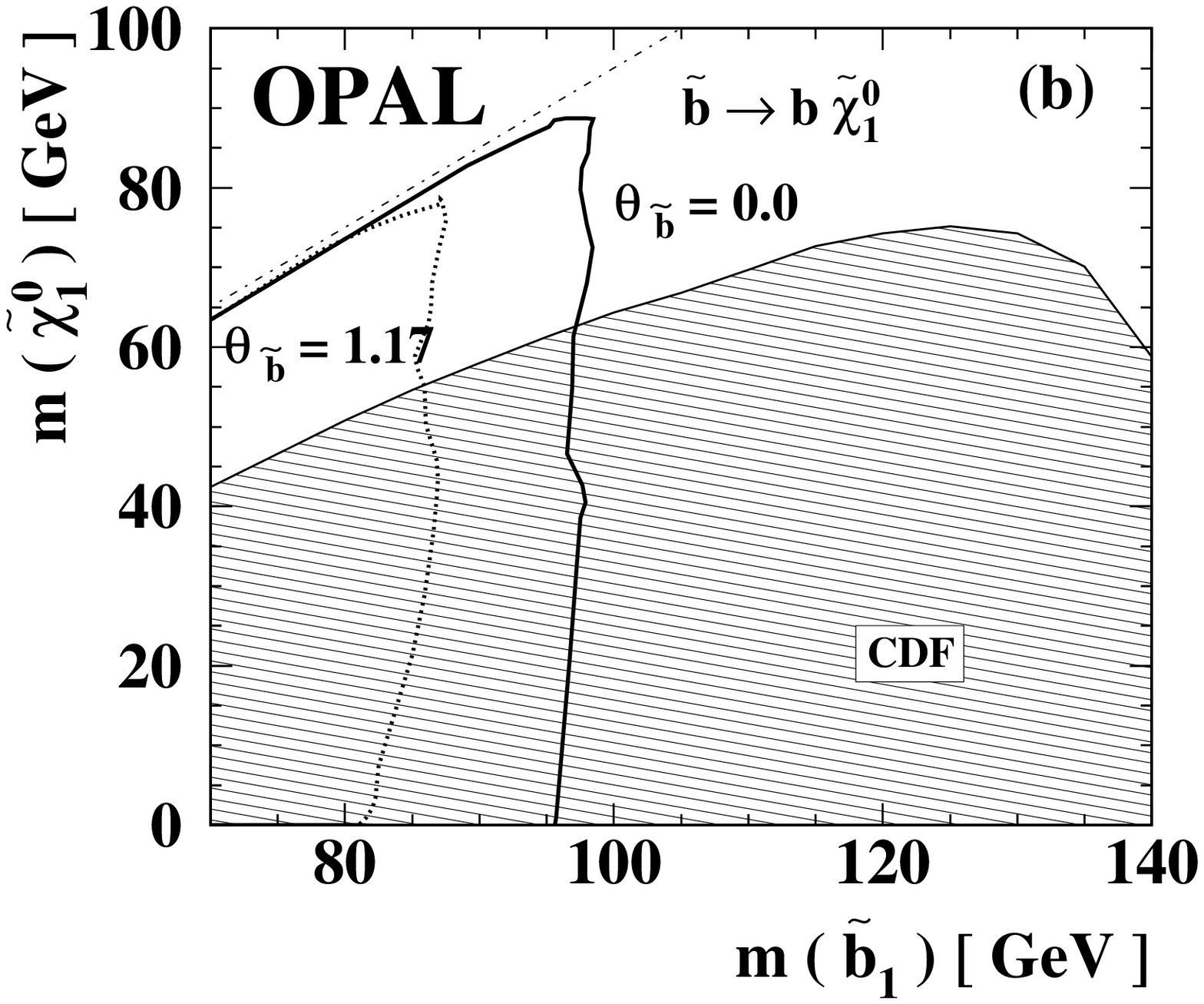,width=12.0cm} 
}\end{center} 
\vspace*{-2.0mm}
\caption[]
{
(a) The 95\% C.L. excluded regions in the $(\mstop, \, \mchi)$ plane
assuming that $\stopm$ decays into $\cq \neutralino$\@.
The solid line shows the limit for
zero mixing angle of $\stopm$, and 
the dotted line shows the limit for 
a mixing angle of 0.98~rad ($\stopm$ decouples from the $\Zboson$ boson)\@.
The dash-dotted straight line shows the kinematic limit 
for the $\stopm \ra \cq \neutralino$ decay.
In the triangular region of $m_{\stopm}-\mchi > m_{\Wpm} + \mb$,
the decay $\stopm \ra \bq \neutralino \Wp$(on shell)
through a virtual chargino becomes dominant.
This region is not excluded. 
(b) The 95\% C.L. excluded regions in the $(m_{\sbotm}, \, \mchi)$ plane,
assuming 
that $\sbotm$ decays into $\bq \neutralino$\@.
The solid line shows the limit
where the mixing angle of $\sbotm$ is assumed to be zero,
and the dotted line shows the limits for a mixing angle of 1.17~rad
($\sbotm$ decouples from the $\Zboson$ boson)\@.
The singly-hatched regions in (a) and (b) are excluded 
by the CDF Collaboration~\cite{CDF}.
}
\label{fig:result1}
\end{figure}
\newpage 
\begin{figure}[htb]
\vspace*{-15.0mm}
\begin{center}\mbox{
\epsfig{file=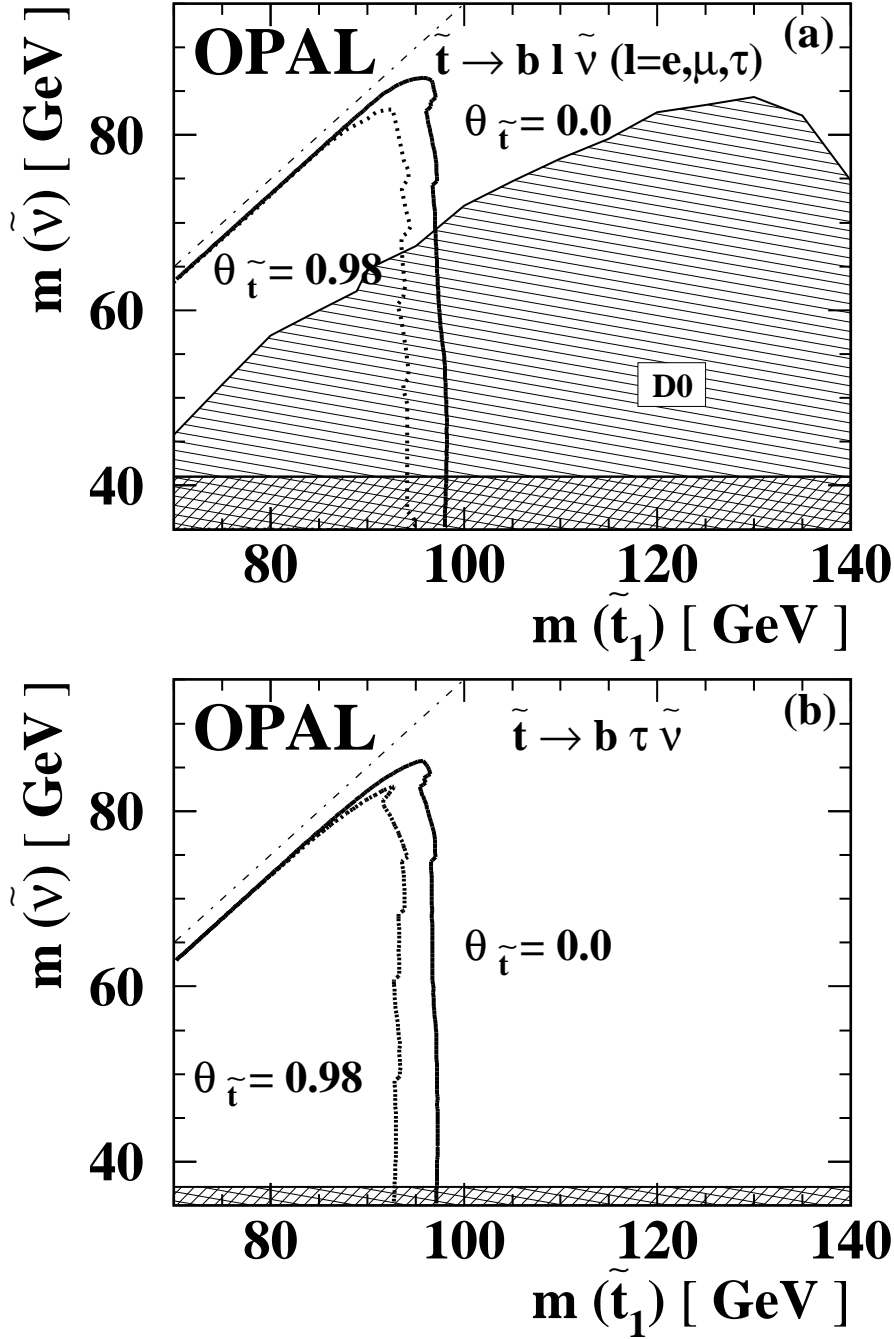,width=14.0cm}
}\end{center} 
\vspace*{-2.0mm}
\caption[]
{
The 95\% C.L. excluded regions in the $(\mstop, \, m_{\snu})$ plane
assuming that the $\stopm$ decays into $\bq \ell \snu$;
(a) the branching fraction to each lepton flavour
is the same;  
(b) $\stopm$ always decays into $\bq \tau \snu_{\tau}$\@.
The solid lines show the limits where the mixing angle of $\stopm$
is assumed to be zero, and 
the dotted lines show the limits for a mixing angle of 0.98~rad
(decoupling case)\@.
The cross-hatched region has been excluded by 
measurements of the $\Zboson$ invisible decay width at LEP1~\cite{snulimit},
and the dash-dotted diagonal line shows the kinematic limit 
for the $\stopm \ra \bq \ell \snu$ decay.
The singly-hatched region in (a) is excluded 
by the D0 Collaboration~\cite{D0}.
 
}
\label{fig:result2}
\end{figure}

\end{document}